\documentclass[preprint,12pt, 1p]{elsarticle}

\usepackage{amssymb}

\usepackage{graphicx}
\graphicspath{
	{images/}
}

\usepackage{subfiles}

\usepackage{upgreek}

\usepackage[hyphens]{url}

\usepackage{amsmath}
\usepackage{siunitx}

\journal{arXiv}

\begin{document}

\begin{frontmatter}



\title{An Analysis of Optical Contributions to a Photo-Sensor's Ballistic Fingerprints\tnoteref{copyright}}


\author[eee]{R.~Matthews\corref{cor1}}
\ead{richard.matthews@adelaide.edu.au}
\author[eee]{M.~Sorell}
\author[cs]{N.~Falkner}
\cortext[cor1]{Corresponding author}
\tnotetext[copyright]{ Copyright 2018. This manuscript version is made available under the CC-BY-NC-ND 4.0 license http://creativecommons.org/licenses/by-nc-nd/4.0/}

\address[eee]{The University of Adelaide, School of Electrical and Electronic Engineering, Adelaide, SA, 5005 AUS.}
\address[cs]{The University of Adelaide, School of Computer Science, Adelaide, SA, 5005 AUS.}

\begin{abstract}
Lens aberrations have previously been used to determine the provenance of an image. However, this is not necessarily unique to an image sensor, as lens systems are often interchanged. Photo-response non-uniformity noise was proposed in 2005 by Luk\'{a}\v{s}, Goljan and Fridrich as a stochastic signal which describes a sensor uniquely, akin to a ``ballistic'' fingerprint. This method, however, did not account for additional sources of bias such as lens artefacts and temperature.

In this paper, we propose a new additive signal model to account for artefacts previously thought to have been isolated from the ballistic fingerprint. Our proposed model separates sensor level artefacts from the lens optical system and thus accounts for lens aberrations previously thought to be filtered out. Specifically, we apply standard image processing theory, an understanding of frequency properties relating to the physics of light and temperature response of sensor dark current to classify artefacts. This model enables us to isolate and account for bias from the lens optical system and temperature within the current model. 

\end{abstract}


\end{frontmatter}



	\section{Introduction}

	Much work has been done to solve the open and closed set camera identification problem \cite{lukas2006digital, alles2009source, knight2009analysis, geradts2001methods, dirik2008digital}. One of the most promising methods used to identify an image uniquely to not just a particular make or model of camera but the unique image sensor itself is that of photo-response non-uniformity noise (PRNU) or sensor pattern noise (SPN) \cite{lukas2006digital}. While blind source camera identification has been used for some time as a reliable and accepted method for legal purposes \cite{strachan2009}, there are untested scenarios within the existing literature that provide a level of uncertainty. It is widely accepted best practice to identify any source of error within forensic tools and provide methods for their mitigation \cite{SWGDE_ErrorMit}. There still remains questions regarding the science of the method due to high-frequency components of the image remaining within either the image fingerprint, the camera reference pattern or both. These high frequency components include but are not limited to image compression artefacts \cite{alles2009source}, dark current \cite{holst2007cmos}, amplifier noise \cite{holst2007cmos}, and lens and optical effects \cite{knight2009analysis} including dust \cite{dirik2008digital}. These high frequency components can corrupt the fingerprint if their energy dominates the unique signal and are significantly uncorrelated to the sensor. In this paper we aim to isolate a source of error from blind source camera identification and, applying principles of signal processing, demonstrate the energy distribution to the various traces that the SPN method is capable of isolating.

	While much is known about the mathematical design of lenses, only recently have image analysts begun to study their unique geometric effects to solve the camera identity problem \cite{san2006source, johnson2006exposing}. Lens aberrations have successfully been used in image linking \cite{san2006source} and identifying copy paste forgeries \cite{johnson2006exposing}. This is because lenses create artefacts in an image known as Seidel Aberrations \cite{Seidel_1857}. These aberrations describe how each ray of light travelling through a lens deviates in some manner from the optical axis and is unique to a lens system due to the multiple lenses used in combination \cite{jenkins1937fundamentals}. While this method is successful at lens identification, it provides little information about the specific image sensor in question since lens systems are easily substituted. 
	
	\begin{figure*}[!t] 
		\centering
		\includegraphics[width=0.8\textwidth]{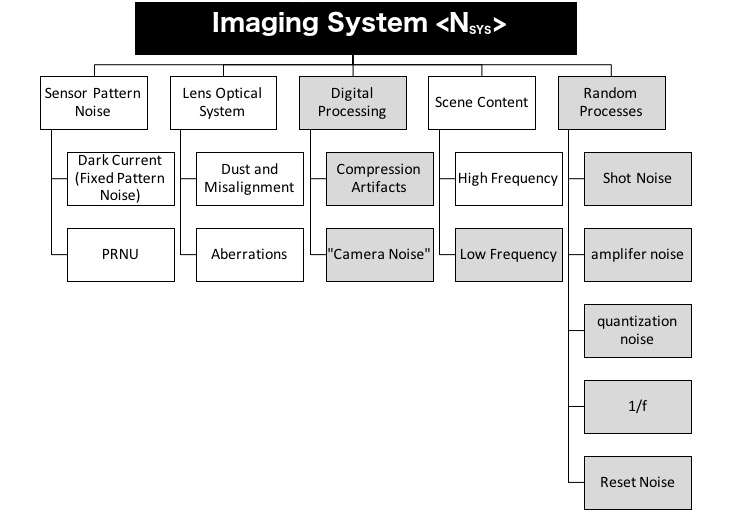}
		\caption{The noise residue model as proposed in our work based on the system noise equations from \cite{holst2007cmos}. The dark grey boxes indicate sources of noise that can be easily mitigated. Random processes are mitigated through frame averaging \cite{holst2007cmos} while RAW format images remove digital processing artefacts\cite{knight2009analysis}. Camera Noises introduced through computer algorithms such as demosaicking, dynamic range adjustments and downsampling are controlled through correct choice of camera hardware values prior to imaging \cite{holst2007cmos}. The low-frequency components of the scene content and all other sources of noise are removed due to the high-pass filter that the images are passed through to obtain the noise residue.}
		\label{newModel-isolation}
	\end{figure*}
	
	While an abstract model of noise within image sensors has been developed as seen in Figure \ref{newModel-isolation}, to determine which source dominates a complete signal-to-noise ratio analysis must be undertaken \cite{holst2007cmos}. In this paper, we begin this work by limiting ourselves to evaluating the contribution of the lens within a noise residue to ascertain if contamination is possible with lens substitution. The next section provides an overview of how the noise residue is obtained and describes the work that has already been done in isolating the contaminating effects within this fingerprint. In Section \ref{sec:background} we describe a new noise model for the noise residue based off the work of \cite{holst2007cmos} that is more inclusive of the high-frequency leakage seen in \cite{lukas2006digital}. We describe our lens isolation experiments in Section \ref{sec:experimentsetup}  in which we use a physical filter to remove all contributions from Seidel aberrations. The results of these experiments are discussed in Section \ref{sec:resultsanddiscussion} before concluding in Section \ref{sec:conclusion}.

	In the sections which follow we consider all operations as element-wise matrix operations unless specifically expressed otherwise. Boldface is used to denote $m\times n$ vectors. The product between two vectors is assumed to be the vector product $ \mathbf{x}\odot \mathbf{y} = \sum^n_{i=1} \mathbf{X}[i]\mathbf{Y}[i] $ where $i$ is the i'th element of the vector. $||\mathbf{X}|| = \sqrt{\mathbf{X}\odot <{X}>}$ is used to denote the argument of the vector $\mathbf{X}$,  and the mean value of the vector $\mathbf{X}$ is indicated by$\mathbf{<{X}>}$. Correlation between two vectors is the cross correlation:
	
	\begin{displaymath}
	corr(X,Y) = \frac{(\mathbf{X}-{\mathbf{<{X}>}})\odot (\mathbf{Y}-{\mathbf{<{Y}>}})}{||\mathbf{X}- {\mathbf{<{X}>}}||\odot||\mathbf{Y}-{\mathbf{<{Y}>}}||} \nonumber
	\end{displaymath}
	
	\section{Background}\label{sec:background}
	
\begin{figure}[!t]
		\centering
		\includegraphics[width=21pc]{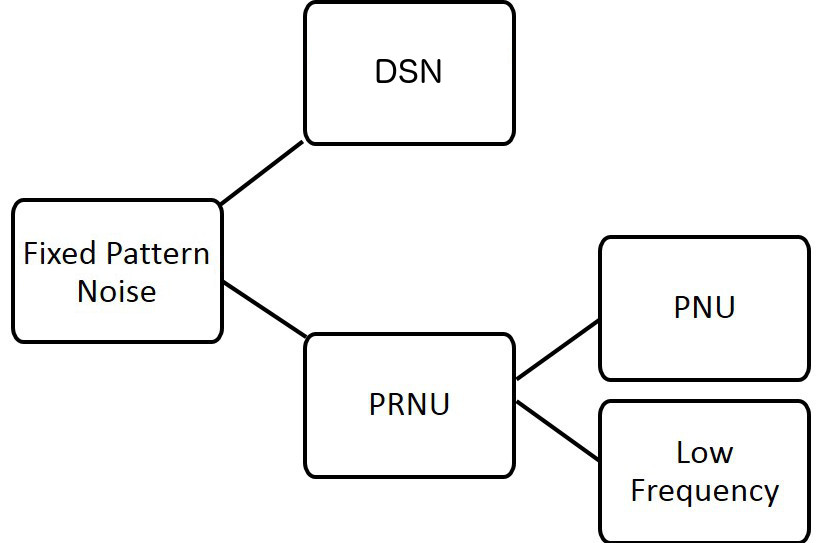}
		\caption[Luk\'{a}\v{s} \emph{et al} noise model]{The additive noise model as proposed by Luk\'{a}\v{s} \emph{et al} in their seminal work\cite{lukas2006digital}}
		\label{oldModel}
	\end{figure}

	We model the signals contained within a single image as an additive signal model based on \cite{holst2007cmos} and shown in Figure \ref{newModel-isolation}. This expands upon the model proposed by \cite{lukas2006digital} as shown in Figure \ref{oldModel}.
	
	Quoting levels of noise in terms of electrons at the level of image sensor output, the noise magnitude is the root mean square value and the sources are expressed as the root sum of the squares and added in quadrature where appropriate we obtain the following:

	\begin{align*}
	\begin{split}
	<\mathbf{N_{SYS}}>= &\sqrt{<\mathbf{n^2_1}> + ... + <\mathbf{n^2_i}>} \\ &\overline{ + ... + <\mathbf{n^2_N}>}
	\end{split}
	\end{align*}

	Where $ <\mathbf{n^2_i}> $ is the variance of noise source $ i $ and $ <\mathbf{N_{SYS}}> $ is the standard deviation measured in RMS units for the entire system.

	Substituting for the various sources of noise identified in Figure \ref{newModel-isolation} :
	
	\begin{align}
	\begin{split}
	<\mathbf{N^2_{SYS}}{>} {\medspace} {=} &<\mathbf{n^2_{RANDOM}}> + <\mathbf{n^2_{IMAGE}}> \\
	{\medspace}&{+}\: <\mathbf{n^2_{DIGITAL}}>  + <\mathbf{n^2_{LOS}}>  \\
	{\medspace}&{+}\: <\mathbf{n^2_{SPN}}>
	\end{split}
	\end{align}
	
	Since SPN is the signal we wish to isolate we deviate from traditional noise models to include the image as a noise source where $\mathbf{n_{IMAGE}}$ is the high and low frequency components of the scene being imaged, $\mathbf{n_{DIGITAL}}$ are the noise sources due to the digital processing pipeline, $\mathbf{n_{LOS}}$ is the Lens Optical System (LOS) , $\mathbf{n_{SPN}}$ is the contribution from SPN being the addition of dark current (FPN) and PRNU:
	
	\begin{align}\label{2}
	\begin{split}
	<\mathbf{n^2_{SPN}}>{\medspace} {=} &<\mathbf{n^2_{FPN}}> + <\mathbf{n^2_{PRNU}}>
	\end{split}
	\end{align}
	
	and $\mathbf{n_{RANDOM}}$ is the sources of noise able to be mitigated through frame averaging represented by:
	
	\begin{align}\label{3}
	\begin{split}
	<\mathbf{n^2_{RANDOM}}{>} {\medspace} {=} &<\mathbf{n^2_{SHOT}}> + <\mathbf{n^2_{A}}>  \\
	{\medspace}&{+}\: <\mathbf{n^2_{ADC}}> + <\mathbf{n^2_{\frac{1}{f}}}> \\
	{\medspace}&{+}\:<\mathbf{n^2_{RESET}}> 
	\end{split}
	\end{align}

	For our usage, we agree with the findings of \cite{holst2007cmos} as shown in our theoretical model of the noise sources contained within the noise residue after following the de-noising method in \cite{lukas2006digital}.
	
\begin{figure}[!t]
		\centering
		\includegraphics[width=21pc]{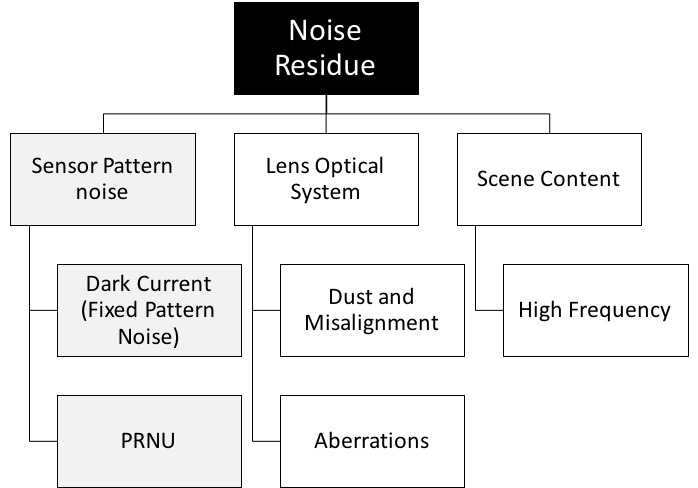}
		\caption[Noise Residue Model]{The groupings of noise remaining in our noise residue. The grey indicates the sensor specific noise.}
		\label{noiseresidue}
	\end{figure}

	We break down the signals within our of noise residue as being comprised of three main areas: SPN or those due to the sensor, those due to the LOS, and the high-frequency components of the scene content. We illustrate this in Figure \ref{noiseresidue}. From the sensor, we follow the model as proposed in \cite{lukas2006digital} and break the Sensor level noise down to PRNU and Dark Current. For ease of modelling, we also include dust on the sensor as per \cite{lukas2006digital} noting that dust modifies the PRNU response since light is blocked from the sensor. The LOS is comprised of two levels. These are lens dust or misalignment, and Seidel aberrations cause by design errors during lens manufacture. Both aspects are high-frequency components only due to the filter $f$ that the system is run through to obtain the noise residue.
	
	To simplify the model, acknowledging we introduce a source of possible error in doing so, we focus our attention on the sections of the model that positively correlate with an individual image under test (IUT).  \cite{lukas2006digital} using \cite{holst2007cmos} identified that the only sources of noise not reduced through frame averaging were dark current and PRNU. This was further refined in \cite{knight2009analysis} to eliminate compression level artefacts through the use of raw images. LOS aberrations and components of the scene remain (Figure \ref{noiseresidue}). The scene components are limited to only the spatial high-frequency components since the image has been high-pass filtered. The model when a reference pattern is compared to a IUT fingerprint is therefore considered as follows:
	
	\begin{align}\label{4}
	\begin{split}
	<\mathbf{N^2_{SYS}}{>} {\medspace} {=} &<\mathbf{n^2_{SPN}}> + <\mathbf{n^2_{LOS}}> \\
	{\medspace}&{+}\: <\mathbf{n^2_{W_{ref}}}>  + <\mathbf{n^2_{w}}>  \\
	\end{split}
	\end{align}

	Where $\mathbf{n_{W_{ref}}}$ is the contribution of high-frequency elements from the reference pattern due to the insufficient suppression from frame averaging and $\mathbf{n_{w}}$ is the high-frequency scene components of the IUT.  Given that these two sources are uncorrelated we further reduce our model to:
	
	\begin{equation}\label{14}
	<\mathbf{N_{SYS}}^2{>} {\:} {=} <\mathbf{n^2_{SPN}}> + <\mathbf{n^2_{LOS}}>+<\mathbf{n^2_{V}}> 
	\end{equation}

	From this model, we can determine the contribution of the LOS aberrations within the system to the correlation energy $<\mathbf{N^2_{SYS}}>$. We achieve this through the use of a physical filter (a pinhole lens) thus removing LOS aberrations from the system altogether.
	
	\begin{equation}\label{15}
	<\mathbf{N^2_{SYS}}{>} {\:} {=} <\mathbf{n^2_{SPN}}> +<\mathbf{n^2_{V}}> 
	\end{equation}

	Given $<\mathbf{n_V}>$ is uncorrelated, the resulting correlation will be directly proportional to a combination of FPN and PRNU. We assume the definition of this as SPN as per \cite{holst2007cmos}. This is the basis of the original method as seen in \cite{lukas2006digital} with differences as explained here. \cite{lukas2006digital} acknowledged that pattern noise is \emph{any noise component that survives frame averaging} and focused on only one part of this theoretical model, pixel non-uniformity noise (PNU). From our use of the model as proposed in \cite{holst2007cmos} it is clear that lens aberrations are involved unless otherwise filtered, which frame averaging does not achieve. \cite{lukas2006digital} assumed the positive match from their method was directly proportional to PNU defined as a sub layer of PRNU caused by the different sensitivity of pixels to light. We do not agree as shown from the theoretical break down above hypothesis that Dark Current and the LOS must be included with appropriate weight.
	
	Dark current is generated in multiple places within an image sensor. Generally, three sources of dark current contribute to the total generated by a sensor. These sources are typically the depletion region through the swapping of minority carriers, the diffusion of carriers in the field-free region at saturation (drift current) and on the surface of any oxide layer interface. A complete study of dark current is beyond the scope of this paper but can be read in \cite{widenhorn2002temperature}.  Previous work by \cite{kurosawa1999ccd} has presented a hypothesis that dark current could be an effective tool for matching images to a source camera. However, this work is often reduced to pixel defects for matching images as demonstrated in \cite{geradts2001methods}. We propose that even with pixel defects isolated and removed, dark current is a unique trace in itself. This philosophy has previously been proposed. In \cite{kurosawa2013casestudies} the concept of a hybrid model was explored were the individual traces of PRNU and DSN were combined to create a new method that ultimately ``had higher discrimination capability than the method using pixel non-uniformity when the number of recorded image was small''. By isolating lens effects and dark current from the PRNU trace we illustrate why this is the case further demonstrating the need for more work to understand the science behind the sensor pattern method for uniquely solving the blind source camera identification problem.
	
	\section{Methods}\label{sec:experimentsetup}
	
	Six Sony IMX219PQ CMOS image sensors (CIS) with integrated lenses were used in our analysis. The lenses were carefully removed from the sensor and placed into a 3D printed jig. The jig was designed explicitly so that each image was slightly out of focus. This assists in removing high-frequency image components from our analysis. Only three sensors were used in our experiments as three sensors were damaged during the lens removal process. This gave us six lenses and three image sensors. Images were taken of a fixed intensity white LED light source evenly focused through a sheet of white opaque perspex to create an evenly illuminated light box. This suppressed contamination from high-frequency image content being passed through the high-pass filter of our PRNU estimator. 
	
	Pinholes were manufactured using 3D printing. A 1.5mm diameter aperture was designed at a distance of 3mm to ensure the focal ratio of the lens was kept consistent at f/2.0. This enabled intensity of the light striking the sensor to be keep consistent across the pinhole and standard lenses resulting in a consistent integration time of 1/1008 seconds. Ensuring integration time was consistent meant that no scaling effects occur between pinhole and lens image sets, and keeps dark current constant. The temperature was kept at a constant T=\SI{30}{\celsius} to further ensure the effects of dark current were controlled. Varying either the exposure light intensity of the temperature should result in an increase of dark current, hence, it is important that these variables are kept constant for comparison.
	
\begin{figure}[!t]
		\centering
		\includegraphics[width=21pc]{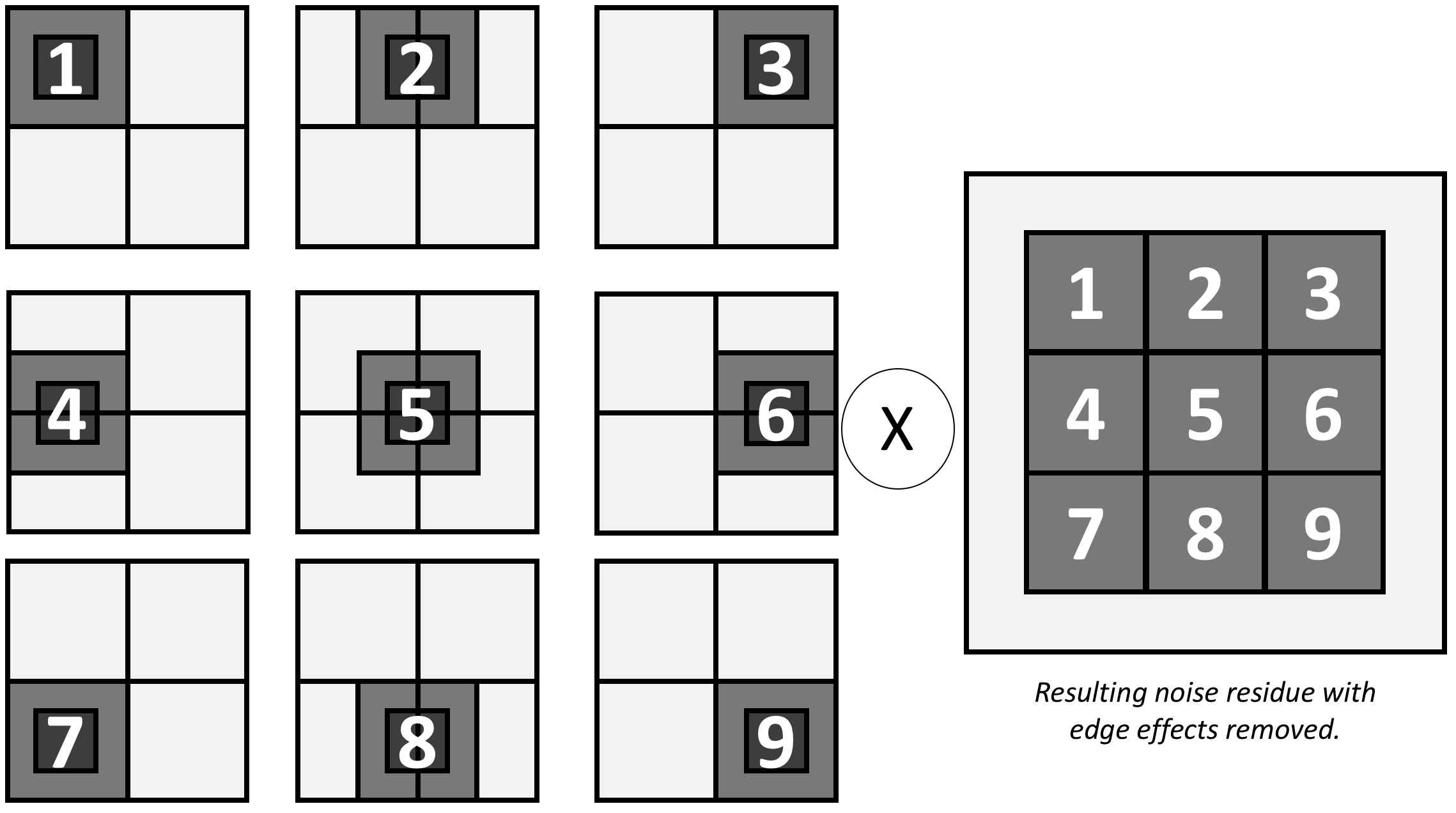}
		\caption[Overlapping Edge Removal Method]{The overlapping method to reduce edge effects from the wavelet coring method. The dark regions in each square (left) indicate the calculated region of the filtered noise residue retained in each pass corresponding to their effecting location in the resulting noise residue (right). The method results in m+1 x n+1 passes being performed as opposed to the original m x n.}
		\label{newAlgorithm}
	\end{figure}

	Each image was preprocessed before filtering. We separated each colour filter array element into a separate array. Each image was cropped to 1024x1024 image offset by 38 pixels from the top left-hand corner. This enabled us to obtain a broad cross-section of the image and would emphasis any lens effects such as vignetting. The resulting four arrays were turned into zero mean signals before being processed by the wavelet coring filter \cite{farid2016photo}. We used the same wavelet coring filter as proposed in \cite{lukas2006digital} with one minor difference. Instead of using advancing squares in the m x n MAP estimate we used overlapping squares, doubling the number of calculations required. We then rejected the outer edges of the m x n pixel arrays to ensures edge effects are discounted from the final analysis. The m x n arrays are then stitched back to create the final PRNU analysis for each CFA array. This process is shown in Figure \ref{newAlgorithm}. Finally, we merge each PRNU estimation for each CFA array back to a single array for the PRNU estimation of the entire sensor. We then correlate each CFA to its corresponding CFA in the image under test. Our final correlation value is then taken to be the average correlation value across each of the four CFA sub arrays. 
	
	150 images were taken using each of the three cameras with each of the six lenses attached in turn. From each set of 150 images, 50 were randomly divided into a reference pattern, while the remaining 100 formed an image set to correlate against the reference patterns. An additional set of images was captured at the same exposure time, temperature and illumination using a pinhole designed to have the same f-number as the original lenses. This gave us seven sets of images per camera and 21 discrete reference patterns. 
	
	\section{Results and Discussion}\label{sec:resultsanddiscussion}
	
\begin{figure}[!t]
		\centering
		\includegraphics[width=21pc]{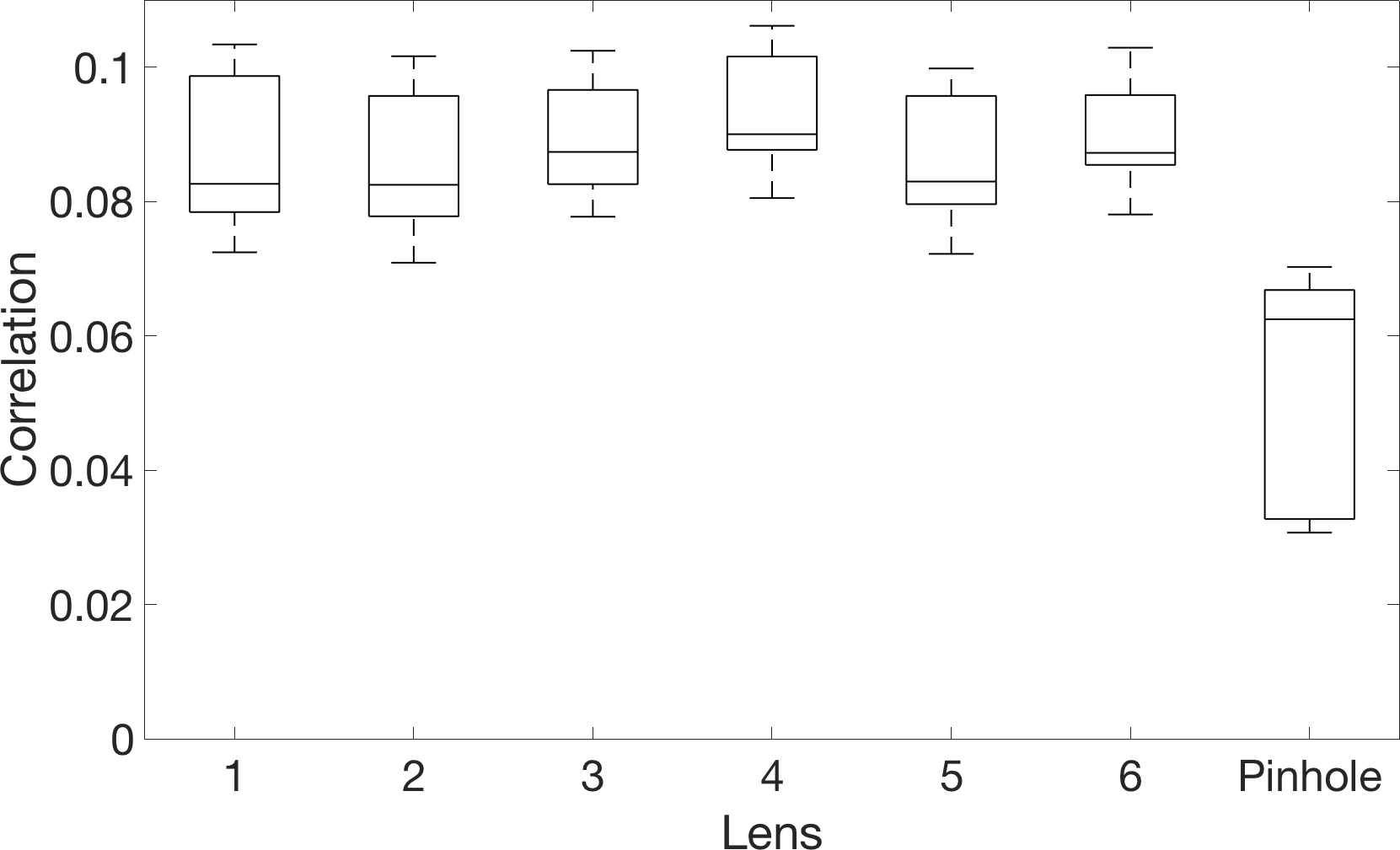}
		\caption[All Cameras with Dark Current]{Box plot of all cameras reference patterns of the lenses correlated against lens image sets not corrected for dark current.}
		\label{allcameraswith}
	\end{figure}

\begin{figure}[!t]
		\centering
		\includegraphics[width=21pc]{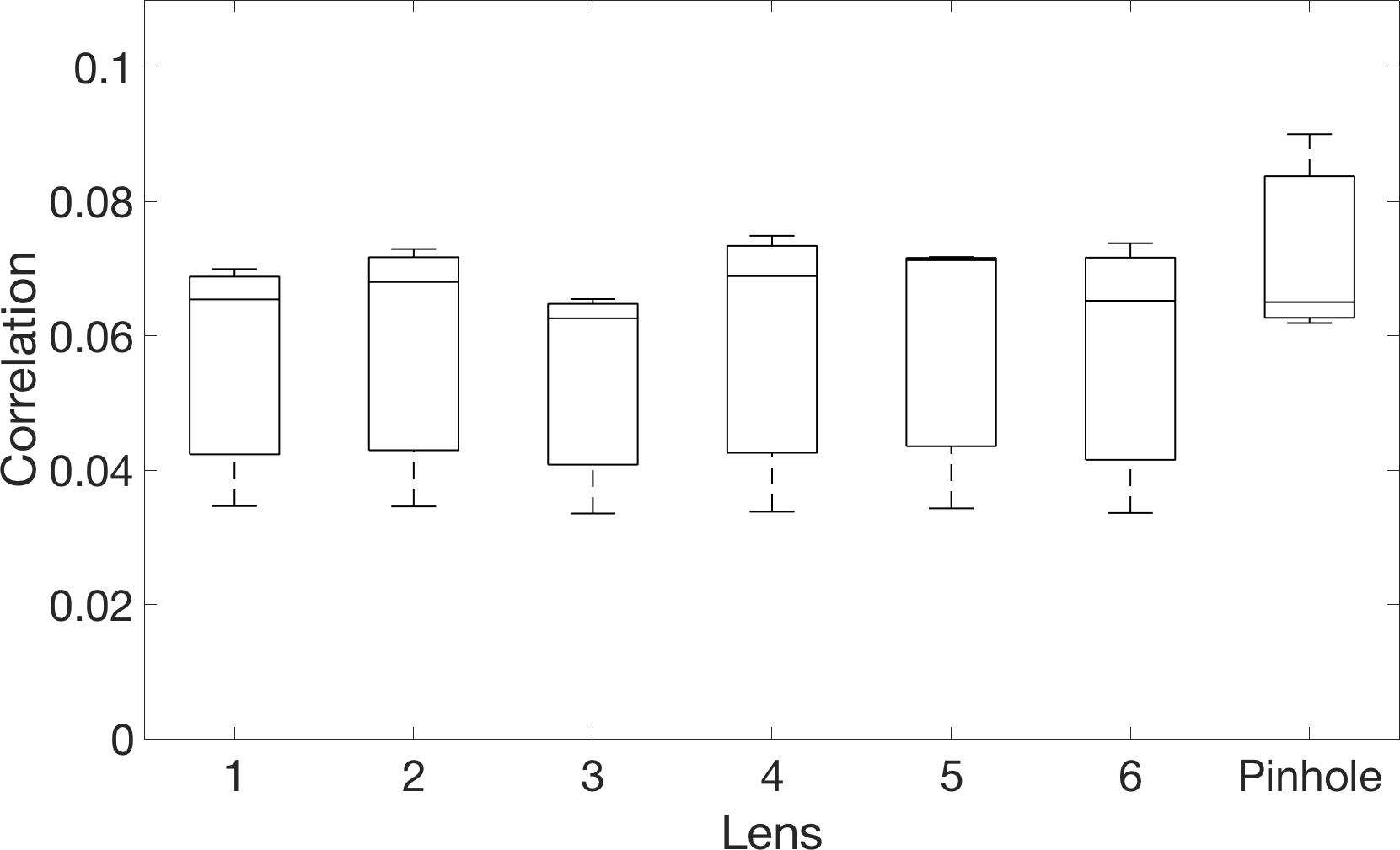}
		\caption[Pinhole Image Sets with Dark Current]{Box plot of pinhole image reference patterns correlated against pinhole image sets not corrected for dark current.}
		\label{pinholewith}
	\end{figure}

	Results of the lens image sets (3600 images) correlated against each of the seven reference patterns are shown in Figure \ref{allcameraswith}. Figure \ref{pinholewith} shows the result of these same reference patterns correlated with the remainder images captured using the pinhole lens on each camera (300 images). Only images known to be from that camera are shown in these figures as uncorrelated results are uniformly distributed about the origin and hence are omitted for clarity. 
	
	The results in Figure \ref{allcameraswith} show the lens reference patterns with similar means and ranges. Our results are approximately 0.02 larger than those first reported in \cite{lukas2006digital} which we attribute to the additional steps taken to eliminate edge effects in our denoising filter. Each of the lens sets shows statistically similar results. The range, mean, maximum and minimum values are consistent within an overall range of 0.025 to 0.031 across the six lenses. The pinhole set, however, has a range of almost twice that at 0.040 with the maximum value below the minimum value of any one lens. This suggests statistical invariability across lenses manufactured of the same type, however, reinforces the hypothesis that high-frequency lens artefacts are included in the noise residues used to create both individual fingerprint and reference patterns.


	The lens sets have a median value of 0.085 and mode of 0.083. There is a difference of less than 0.007 within the means for the lens sets showing that they are consistent. The pinhole set has a mean of only 0.062. This is a clear difference between the average means of the lens sets and the mean of the pinhole set at 0.023, but the pinhole is still capable of statistically matching to the right camera. The pinhole correlation showed a broader range than each of the lens sets with a majority of the data falling within the interquartile range and skewed towards the higher values, whereas each of the lens sets is skewed towards the lower end of the range. 
	
\begin{figure}[!t]
		\centering
		\includegraphics[width=21pc]{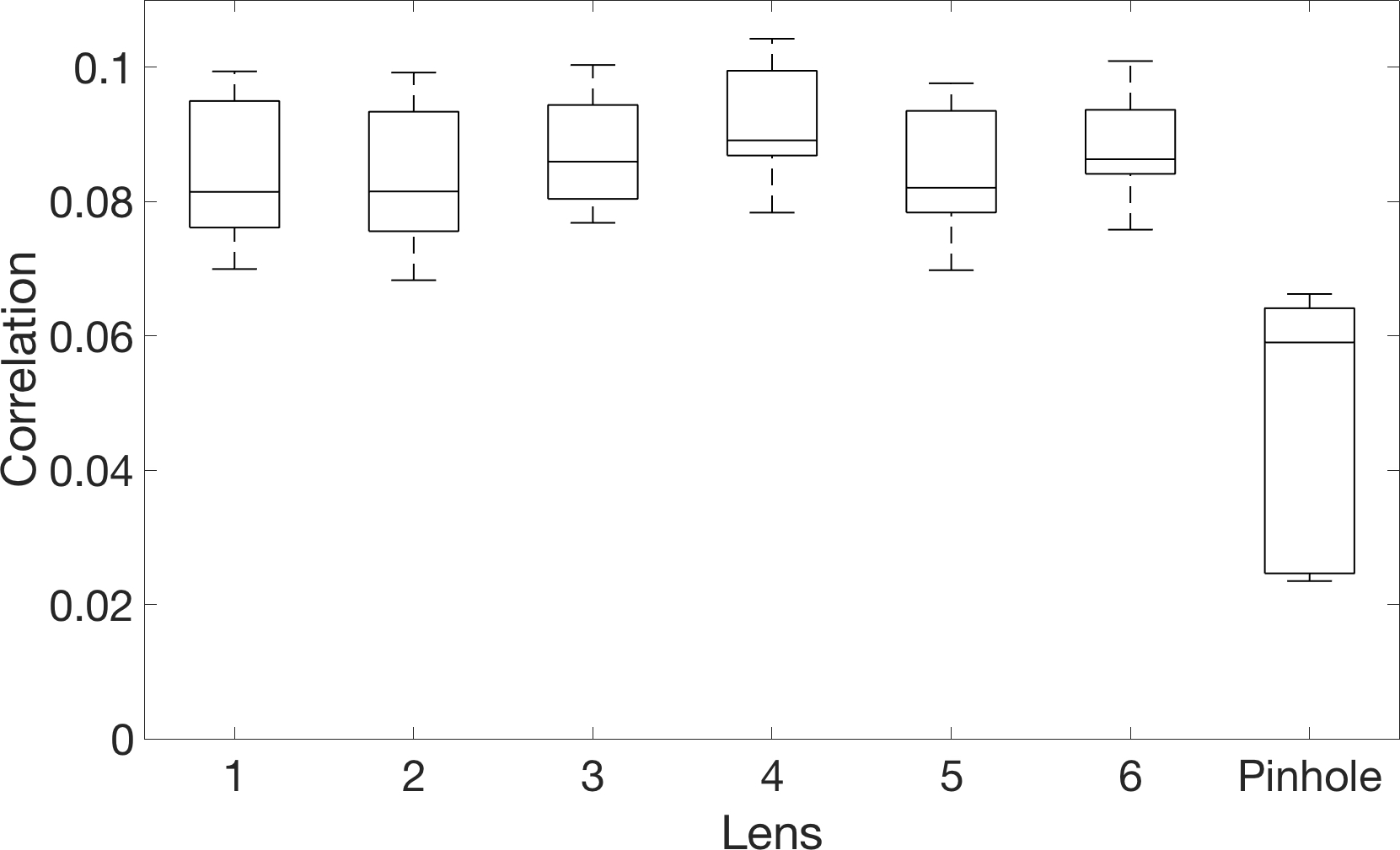}
		\caption[All Camers without Dark Current]{Box plot of all camera reference patterns of the lenses correlated against lens image sets corrected for dark current.}
		\label{allcameraswithout}
	\end{figure}

\begin{figure}[!t]
		\centering
		\includegraphics[width=21pc]{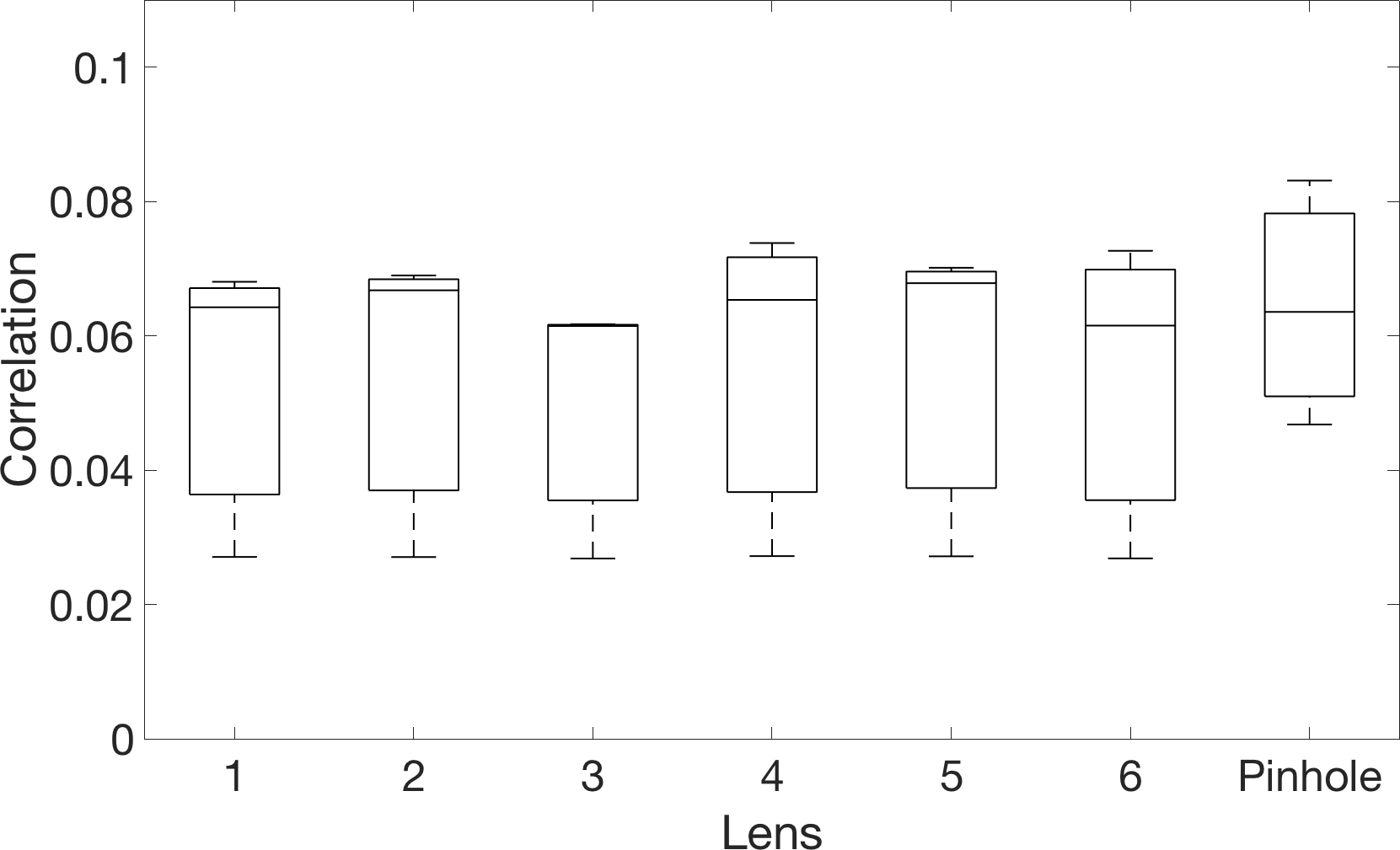}
		\caption[Pinhole Image Sets without Dark Current]{Box plot of pinhole reference patterns correlated against pinhole image sets corrected for dark current.}
		\label{pinholewithout}
	\end{figure}

	An interesting observation is that the uncorrelated pinhole image set is positively skewed whereas the correlated lens image sets are negatively skewed. This observation extends to figures \ref{pinholewith} and \ref{allcameraswithout} with the exception being the pinhole image set matched to a pinhole reference pattern corrected for dark current with dark frame removal in Figure \ref{pinholewithout}.  It may be possible to identify or profile a device such as a pinhole by comparing means of statistically significant sample sizes in this manner.
	
	Comparing Figure \ref{allcameraswith} to Figure \ref{pinholewith} it is apparent that the means of the lens reference patterns reduce to be in line with the pinhole reference pattern when correlated against pinhole image sets taken from the same camera. This aligns with the hypothesis that the lens reference patterns contain additional signal energy from the high-frequency components of the lens passing through the signal filters in the process of obtaining the noise residues.

	Using the average correlation from the lens sets in Figure \ref{allcameraswith} (since they are consistent) and Equation \ref{14} above we can calculate the overall correlation energy of the SPN with the effects of the lens included.
	
	\begin{align}\label{darkcurrent1}
	\begin{split}
	<\mathbf{N^2_{SYS}}{>} {\:} -<\mathbf{n^2_{V}}> & {=} <\mathbf{n^2_{SPN}}> + <\mathbf{n^2_{LOS}}>  \\
	&= 0.0865 
	\end{split}
	\end{align}
	
	Using the averages of all results contained in Figure \ref{pinholewith} we are able to calculate the correlation energy of the SPN without effects of the lens present:
	
	\begin{align}\label{darkcurrent2}
	\begin{split}
	<\mathbf{N^2_{SYS}}{>} {\:} -<\mathbf{n^2_{V}}> -  <\mathbf{n^2_{LOS}}> &{=} <\mathbf{n^2_{SPN}}>  \\
	&= 0.0666
	\end{split}
	\end{align}
	
	Substituting this result back into \ref{14} we can obtain a result for the correlation energy of the LOS alone.
	
	\begin{align}\label{darkcurrent3}
	\begin{split}
	0.0666 + <\mathbf{n^2_{LOS}}> &= 0.0865 \\
	<\mathbf{n^2_{LOS}}>&= 0.0199 
	\end{split}
	\end{align}
	
	This figure corresponds to the effects of the LOS based on measurements with dark current influence in the sensor.

	Since many modern-day camera correct for dark current, each of the images was also corrected for dark current through the use of a dark current frame removal. This was to ensure dark current was not contaminating our results. As seen in Figures \ref{allcameraswithout} and \ref{pinholewithout} the range of the correlation scores are reduced however the overall result remains. The correlation is significantly reduced upon removal of the lens. 
	
	The pinhole set with dark current removal shows good uniformity with the mean centred about the range of the set. Since lens effects and dark current have been removed, the correlation should be acting upon only the correlated PRNU within the image. This shows a Gaussian distribution as expected with the mean centred around the mean of the lens results. When our reference pattern is constructed only with pinhole images and is correlated with images taken using a lens we are no longer correlating against the high-frequency lens artefacts seen within the lens reference patterns. This is a clear result from Figures \ref{pinholewith} and \ref{pinholewithout}.

	We can use this result to further estimate the effects of dark current within the correlation energy by repeating the calculations above and also check the figure we have obtained for the LOS.
	
	Using the average correlation from the lens sets in Figure \ref{allcameraswithout} (since they are consistent) and Equation \ref{14} above we can calculate the overall correlation energy of the PRNU only with the effects of the lens included but without dark current.
	
	\begin{align}\label{nodarkcurrent1}
	\begin{split}
	<\mathbf{N^2_{SYS}}{>} {\:} -<\mathbf{n^2_{V}}>&{=} <\mathbf{n^2_{SPN}}> + <\mathbf{n^2_{LOS}}>  \\
	&= 0.0844  
	\end{split}
	\end{align}
	
	Using the averages of all results contained in Figure \ref{pinholewithout} we are able to calculate the correlation energy of the PRNU alone:
	
	\begin{align}\label{nodarkcurrent2}
	\begin{split}
	<\mathbf{N^2_{SYS}}{>} {\:} -<\mathbf{n^2_{V}}> -  <\mathbf{n^2_{LOS}}> &{=} <\mathbf{n^2_{SPN}}>  \\
	&= 0.0644 
	\end{split}
	\end{align}
	
	Substituting this result back into \ref{14} we can obtain a result for the correlation energy of the LOS alone. 
	
	\begin{align}\label{nodarkcurrent3}
	\begin{split}
	0.0644 + <\mathbf{n^2_{LOS}}> &= 0.0844 \\
	<\mathbf{n^2_{LOS}}>&= 0.0200
	\end{split}
	\end{align}
	
	Figure \ref{pinholewith} corresponds to the effects of the lens only within the sensor and is consistent for our measurements with dark current as calculated in Equation \ref{darkcurrent3}. Comparing the results of Equations \ref{darkcurrent1} and \ref{darkcurrent2} with Equations \ref{nodarkcurrent1} and \ref{nodarkcurrent2} we see that dark current (FPN) corresponds to a contribution of ~0.0022 to a correlation of 0.0865.
	
	Given these values represent power correlation amplitudes we are able to convert them into signal to power ratio terms using the following:
	
	\begin{align}\label{SNR}
	\begin{split}
	SNP_{dB} &= 10 log_{10} \frac{P_{signal}}{P_{total}} \\
	\end{split}
	\end{align}
	
\begin{figure*}[!t]
		\centering
		\includegraphics[width=\textwidth]{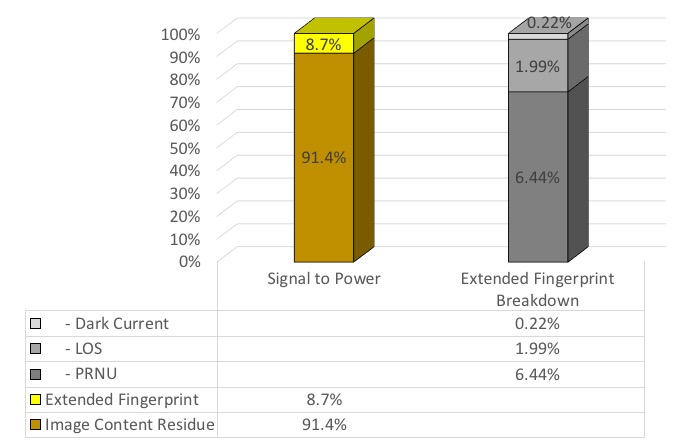}
		\caption{Signal power expressed as a ratio of total power levels of each identifier contained within our extended fingerprint.}
		\label{SNRValues}
	\end{figure*}

	Where the identifier is either PRNU, dark current or combination of them both. These values are summarised in Figure \ref{SNRValues}.
	
	It is possible to treat each of PRNU, dark current and the LOS as unique identifiers hence, the signal. Conversely, we can think of the high frequency image content within our residues as the noise. Switching our definition of the signal and noise in such a manner enables us to calculate SNP values for each identifier in a forensic context. These values are summarised in Figure \ref{SNRValues}. We see that the uncorrelated high-frequency components of the reference patterns dominate with a signal to power value of 91.4\% of the total signal power. PRNU has the strongest of the individual identifiers with a signal to power value of 6.44\% however, the highest identifier value corresponds to the combination of PRNU, dark current and LOS with a value of 8.7\% of the total power of the signal. This is clearly contrary to assumptions made in \cite{lukas2006digital} that the method is matching only to the PNU as a subsection of PRNU since the best match is made with a combination of all the identifiers measured. We can also see that PRNU accounts for nearly 75\% of the extended fingerprint's power. This is compared to the LOS at 23\% and Dark Current at only 2.5\%. This observation provides an explanation as to why the work of \cite{lukas2006digital} was successful although all sources of correlated power were unaccounted for in their method. 
	
	Dark current was measured to be an appreciative 0.2\% of the total range of the image when held constant at T=\SI{30}{\celsius}. This is expected to increase with temperature as seen by the theoretical models in \cite{holst2007cmos} and will be explored in future work. It is also noted that should the intensity of light decrease the amount of power that the dark current has compared to other sources of power in the image would inversely increase. This has not been experimentally shown due to limitations of the lighting apparatus used when conducting this initial study but, we note the theory illustrated in \cite{holst2007cmos}. By removing the lens, we were able to eliminate the source of stochastic variance and isolate the deterministic component of the lens optical system and measure the contribution of the LOS to be 2.0\%. We note that some variance due to lens aberrations will still be present due to the involvement of the micro-lens array on the sensor itself. Some sensors use a dual micro-lens design. These aberrations from the micro-lens are unique to each sensor and hence form a significant source of the PRNU. Likewise, some aspects of the camera noise \cite{holst2007cmos} is unique to each camera but were excluded from our noise residue model since we were only concerned with the image sensor. In reality, since these camera noises are unique their effects may be seen within a sensor fingerprint but have not been attributed above.  
	
	
	
	While cameras of a similar make and model have been evaluated here to eliminate possible sources of experimental error it is worth noting that we expect that other cameras should exhibit similar breakdowns with the amount of power scaling in proportion with the quality of the sensor.  Scientific grade sensors with low dark noise by design should show very little dark current contamination whereas we expect low grade CMOS cameras for integrated mobile applications designed without built in dark current correction to have some of the worse. This expectation is consistent with the results shown here. The Sony IMX219 sensor used in this work has built in dark current correction at the silicon level \cite{sonyIMX219} , however, still shows evidence of a contribution, albeit small, of signal power attributed to this forensic trace.
	
	\section{Conclusion}\label{sec:conclusion}
	
	While we do not dispute that the method first proposed by Luk\'{a}\v{s} \emph{et al} is capable at blindly identifying images uniquely to their source camera, our work has shown that there is more to understand behind this methodology then first described. The additional factors are acting upon the correlation seen need to be understood before it can be used as a reliable methodology to solve the blind source camera identification problem for legal purposes. We have shown that dark current and the LOS contribute a non-insignificant amount of energy to the correlation. While an amount of energy in the correlation is contributed by the lens aberrations, the method shown here is not statistically capable of uniquely identifying a particular lens of similar design. An area of counter forensic method however is left as a proposal to disrupt the SPN fingerprint methods by designing a lens with extreme high-frequency aberrations to corrupt the SPN. This reinforces the initial findings of Knight \textit{et al}. Our method demonstrates an additional protocol step of lens isolation using a pinhole camera should the suspect camera be available to the investigator. Our paper demonstrates that this critical step of image verification should be taken to increase the certainty of a positive match particularly in the context of charges relating relating to the production of photographs using professional grade dSLR cameras where multiple lenses are interchangeable. Such a step can increase the certainty of a positive match and aim to verify the results of the forensic investigation. Forensic investigators must be aware of the result of this step demonstrating the significance that the lens system may play particularly when using SPN to solve the open set blind source camera identification problem.

\section{Acknowledgements}
This research did not receive any specific grant from funding agencies in the public, commercial, or not-for-profit sectors. This work was supported with supercomputing resources provided by the Phoenix HPC service at the University of Adelaide. This research is supported by an Australian Government Research Training Program (RTP) Scholarship and forms part of a thesis chapter. 




  \bibliographystyle{elsarticle-num} 
  \bibliography{TIFS-LensIsolation}

\end{document}